\documentclass[
 prl,twocolumn,amsmath,amssymb,
 aps,superscriptaddress
]{revtex4-2}

\usepackage[colorlinks=true,allcolors=blue]{hyperref}
\usepackage{hyperxmp}
\usepackage{bm}
\usepackage{float}
\usepackage{booktabs,tabularx,array}
\usepackage{graphicx,color}
\usepackage[dvipsnames]{xcolor}

\usepackage{enumerate}
\usepackage{harpoon}
\usepackage{amsmath,amssymb}
\usepackage{gensymb}
\usepackage{accents}
\usepackage[e]{esvect}
\usepackage[type={CC},modifier={by},version={4.0}]{doclicense}
\usepackage{footmisc}

\definecolor{ctcolor}{rgb}{0.8, 0.0, 0.2}

\usepackage{dcolumn}
\usepackage{bm}
\usepackage{dsfont}

\begin{document}
\preprint{APS/123-QED}
\vspace*{15px}

\title{Non-Hermitian Origin of Surface Peregrine Soliton and Its Topological Signatures}
\author{Samit Kumar Gupta}
\email[]{skgupta@ciencias.ulisboa.pt\,, samit.kumar.gupta@gmail.com}
\affiliation{Centro de Física Teórica e Computacional and  Departamento de Física, Faculdade de Ciências, Universidade de Lisboa, Campo Grande 2, Edifício C8, Lisboa 1749-016, Portugal}

\begin{abstract}
A wide range of dynamic wave localization phenomena manifest underlying physical effects in diverse areas of physics which often bear signatures of nontrivial coherent wave structures. A Peregrine soliton draws particular interest because of its space-time localization and the prototypical analogy to extreme waves. We show the emergence of strikingly nontrivial complex wave phenomena in the unbroken and broken regimes of parity-time (PT) symmetry of a class of composite optical media based on a PT variant of the standard nonlinear Schrödinger equation (NLSE). It gives rise to a symmetry-protected self-dual pair of chiral bulk Peregrine and anti-Peregrine solitons. Remarkably, a stable soliton propagates unhindered in the exact PT phase and a surface Peregrine soliton emerges in the broken PT phase at the optical interface between two strongly heterogeneous optical media. We show that such a surface mode emanates from the interplay between the non-Hermitian pseudo-self-induced PT potential and a nonlinearity-dispersion modulation scheme forming a non-Hermitian topological domain wall. A persistent collapse and revival dynamics exists between the pair of Peregrine and anti-Peregrine solitons and the surface Peregrine soliton at the optical interface. The topological signatures of the surface Peregrine soliton are discussed. The results could be interesting at the crossroads of nonlinear optics, non-Hermitian physics, and topological phenomena for controllable wave manipulation in complex optical and scattering media. 
\end{abstract}

\maketitle

\textit{Introduction}.--  Non-Hermitian physics based on PT symmetry \cite{bender1998real, el2018non, feng2017non, ozdemir2019parity, miri2019exceptional, gupta2020parity, li2023exceptional} has witnessed growing interest both in theory and experiments \cite{makris2008beam, guo2009observation, ruter2010observation, regensburger2012parity, el2007theory, longhi2009bloch, chong2011pt, longhi2010pt, ramezani2010unidirectional, alexeeva2012optical, miri2012bragg, sukhorukov2010nonlinear, driben2011stability, suchkov2011solitons, gupta2014solitary} where many intriguing physical effects are enabled by the non-Hermitian degeneracy known as the exceptional point (EP)\cite{miri2019exceptional, ozdemir2019parity, wang2019exceptional,zhou2018optical}. In parallel, notable advances have been made at the intersection between nonlinear optics and PT symmetry \cite{ ramezani2010unidirectional, alexeeva2012optical, miri2012bragg, sukhorukov2010nonlinear, driben2011stability, sarma2014modulation, midya2017waveguides, xia2021nonlinear, suchkov2011solitons, gupta2014solitary, konotop2016nonlinear, suchkov2016nonlinear, midya2017waveguides, zhang2021asymmetric}, many of which are based on paradigmatic nonlinear Schrödinger equation (NLSE)-type systems. Nonlinearity in conjunction with PT symmetry endows a system with a wide range of effects and phenomena that do not have purely dissipative counterparts \cite{konotop2016nonlinear, suchkov2016nonlinear}]. 

An alternative class of highly nonlocal completely integrable NLSE has been proposed in which the standard third-order Kerr nonlinear interaction term $|\psi|^2\psi$ is replaced by its PT symmetric analog $\psi^{*}(z,-x)\psi(z,x) \psi(z,x)$, and thus an effective linear self-induced PT potential $V(z,x)=\psi^{*}(z,-x)\psi(z,x) $ is induced by nonlocal Kerr nonlinearity \cite{ablowitz2013integrable}. Subsequent studies demonstrate intriguing nonlinear wave physics arising from  such  systems, including the existence of simultaneous bright and dark solitons \cite{sarma2014continuous}, dark and anti-dark soliton interaction \cite{li2015dark}, higher-order rational solitons \cite{wen2016dynamics}, interaction in discrete systems \cite{xu2017darboux}, exact solutions and symmetries \cite{sinha2015symmetries}, soliton collision in generic cases \cite{rao2020nonlocal}, and so on. On the other hand, Peregrine soliton (PS) \cite{peregrine1983water,kibler2010peregrine,kibler2012observation} is a limiting case of a wide range of solutions to NLSE, including transversely periodic Akhmediev breathers (ABs) \cite{kibler2010peregrine, dai2014controllable, dudley2009modulation} or longitudinally periodic Kuznetsov-Ma (KM) breathers \cite{kibler2012observation,dai2014controllable}. 

Due to space-time localization, Peregrine soliton has received significant attention \cite{bailung2011observation, Hammani11, PhysRevE.90.062909, CHEN20141228} in particular, for concomitant extreme wave dynamics \cite{Shrira, oppo2013self, zamora2013rogue, zaviyalov2012rogue}. However, they are related to the modulationally unstable continuous wave (CW) background. Their practical realization has become difficult, although some highly careful experiments have been suggested \cite{tikan2017universality}. It is particularly significant in nonlinear non-Hermitian wave systems, where the issue of wave instability becomes pronounced. This calls for harnessing new wave strategies to enhance its stability in complex optical media to see if such coherent nonlinear structures can survive in increasingly generic physical environments. 

\begin{figure*}
    \centering
    \includegraphics[width=0.90\linewidth]{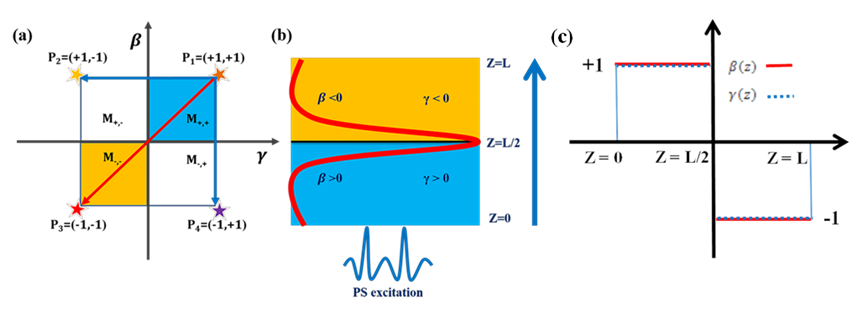}
    \caption{(a) The schematic phase diagram in $\beta-\gamma$ plane that shows the four points ($P_{i}, i=1-4$, denoted by stars) corresponding to the four quadrants of optical media ($M_{+,+}, M_{+,-}, M_{-,-}, M_{-,-}$) depending on the signs of $\beta$ and $\gamma$. The straight red arrow refers to the two partnering optical media with normalized signed values of spatial dispersion and nonlinearity as indicated by points $P_{1}$ and $P_{3}$, whereas the straight blue arrows represent points $P_{2}$ and $P_{4}$. For simplicity, the initial excitation is injected into the partnering medium with point $P_{1}$. (b) Schematic diagram of the spatial dispersion and nonlinearity engineering of the medium. We are mainly interested in the interface between $M_{+,+}$ and $M_{-,-}$ media that exhibits surface modes. The turquoise (yellow) region of the medium indicates normal spatial dispersion and defocusing nonlinearity (anomalous spatial dispersion and focusing nonlinearity). Red curve represents the surface Peregrine soliton mode. (c) The nonlinearity-dispersion engineering scheme for $z\in(0,\,L)$. }
    \label{fig1}
\end{figure*}   

In this connection, nonlinearity or spatial dispersion management has been proposed as a viable means to address this issue \cite{malomed2006soliton}. It is suggested that judicious wave management schemes in nonlinearity or spatial dispersion could play an important role in stabilizing highly localized but unstable wave phenomena. On the other hand, electromagnetic surface states are known to emerge at the interface between two dissimilar media \cite{polo2013electromagnetic, pitarke2006theory, d1988new, takayama2009observation}. Some works have revealed the role of non-Hermiticity and topology in the emergence of surface Maxwell waves \cite{bliokh2019topological}, and the role of topological indices in controlling the nonlinear evolution of extreme waves \cite{marcucci2019topological}. Moreover, several recent works indicate the possibility of hosting topological wave phenomena beyond solid-state physics, such as the topological Kelvin and Yanai modes in geophysical flows \cite{delplace2017topological}, topological interface states in active matter systems \cite{shankar2017topological}, and plasma waves in toroidal geometries \cite{parker2020nontrivial}, among others. Nontrivial topological features can be induced solely by non-Hermiticity \cite{gu2021controlling,gao2020observation,liu2020gain,savoia2014tunneling}, opening a new pathway to complex solitons and surface waves in nontrivial topological systems with added control knobs \cite{bliokh2019topological, komis2023skin, many2024skin}. It could be imperative to explore the non-Hermitian and topological origins of a more generic class of optical surface modes with tailored wave propagation and localization properties. 

To this aim, we consider a nonlocal PT variant of the standard NLSE with Kerr nonlinearity in the form of a PT symmetric pseudo-self-induced potential. We show that suitable nonlinearity and spatial dispersion engineering of such complex optical media can host three intriguing phenomena: i) stable soliton propagation along the strongly heterogeneous media in the exact PT phase, ii) an enhanced surface localization in the broken PT phase at the interface between two distinct optical media with opposite optical properties, iii) existence of a self-dual pair of Peregrine and anti-Peregrine solitons that undergo a periodic collapse and revival dynamics. 


\textit{Theoretical model}.--We consider the scaled nonlocal nonlinear Schrödinger equation where the third-order Kerr nonlinear interaction term is replaced by its PT symmetric counterpart \cite{ablowitz2013integrable,sarma2014continuous} in the form of a PT symmetric pseudo-self-induced potential $V(z,x)=\psi^{*}(z,-x)\psi(z,x)$:
\begin{equation}
    i\frac{\partial{\psi}}{\partial z}+\frac{1}{2}\frac{\partial^2\psi}{\partial x^2}+\psi(z,x)\psi^{*}(z,-x)\psi(z,x)=0.
    \label{ptnlse_standard}
\end{equation}                                                                                
Here, $\psi(z,x)$ is the dimensionless optical field with $\psi(z,-x)$ being its parity conjugate counterpart, where $x$ and $z$ refer to the normalized transverse co-ordinate and propagation distance. The nonlocality in the nonlinear term captures the non-Hermitian feature, and thus nonlinearity and non-Hermiticity are intermingled in a nonlocal way. In contrast to the standard NLSE, the total optical power $P=\int_{-\infty}^{+\infty}{|\psi|^{2}dx}$ satisfies $\frac{dP}{dz}=\int_{-\infty}^{+\infty}|\psi|^{2}(\psi \psi^{*}(z,-x)-\psi^{*}\psi(z,-x)) dx$. Some of the infinite numbers of constants of motion are the quasi-power $Q=\int_{-\infty}^{+\infty}{\psi^{*}(z,-x) \psi(z,x)dx}$ and the Hamiltonian $H=\int_{-\infty}^{+\infty}{(\psi _{x}(z,x) \psi_{x}(z,-x)-\psi^{2}(z,x) {\psi^{*}}^{2}(z,-x))dx}$ \cite{sarma2014continuous}. The potential $V(z,x)$ is self-induced and is dependent on the optical field $\psi^{*}(z,-x)$. However, the transverse shift induces partly external and partly self-induced, and hence a pseudo-self-induced response. In general, Eqn. (1) possesses the solitons on finite background (SFB) solution which reduces to the lowest order rational soliton, \textit{i.e.} the first-order Peregrine soliton:
\begin{equation}
\psi(z,x)=(1-\frac{4(1+2iz)}{1+4x^2+4z^2})e^{iz}.
\label{peregrine]}
\end{equation}

In our model, the effective shift in the transverse coordinate is introduced as $\varepsilon_{\pm}=\varepsilon_{loc}\mp \varepsilon_{tsp}$(where $\varepsilon_{loc}$ denotes the initial locations of the Peregrine solitons and $\varepsilon_{tsp}$ is the transverse shift parameter) and $x\to x\pm\varepsilon_{\pm}$. For simplicity, we keep $\varepsilon_{loc}=\varepsilon$ throughout the paper. On the other hand, if we do not stick to $\beta(z)=1$ and $\gamma(z)=1$ as in Eq. \ref{ptnlse_standard}, and employ nonlinearity and spatial dispersion managements, Eq. \ref{ptnlse_standard} becomes: 
\begin{equation}
    i\frac{\partial{\psi}}{\partial z}+\frac{\beta(z)}{2}\frac{\partial^2\psi}{\partial x^2}+\gamma(z)\,\psi(z,x)\psi^{*}(z,-x)\psi(z,x)=0.
\label{ptnlse_beta_gamma1}
\end{equation}

Here, we consider a nonlinear dispersive medium with a total length of $L$. The whole medium can be thought of comprising two partnering materials with opposite optical properties in nonlinearity and spatial dispersion, one in $z\in(0,L/2)$ with a focusing nonlinearity and anomalous spatial dispersion, and the other in $z\in(L/2,L)$ with a defocusing nonlinearity and normal spatial dispersion. It is schematically shown in Fig. \ref{fig1}(c):
\begin{equation}
    (\beta(z),\,\gamma(z))= 
\begin{cases}
    +1,& \text{if \,} 0<z\leq L/2\\
    -1,& \text{if \,} L/2 <z\leq L.
\end{cases}
\label{beta_gamma1}
\end{equation}

Furthermore, Eq. \ref{ptnlse_beta_gamma2} along with the condition in Eq. \ref{beta_gamma1} could be simply put into a single equation: 
\begin{equation}
    i\frac{\partial{\psi_{\pm}}}{\partial z}+\frac{\beta(z)}{2}\frac{\partial^2\psi_{\pm}}{\partial x^2}+\gamma(z)\,V(z,x)\psi_{\pm}(z,x)=0,
\label{ptnlse_beta_gamma2}
\end{equation}
where, $\beta(z)=1, \gamma(z)=1$ for $\psi_{+}$ and $\beta(z)=-1, \gamma(z)=-1$  for $\psi_{-}$ and $V(z,x)=\psi_{\pm}(z,x)\psi_{\pm}^{*}(z,-x)$.

Here, $\beta(z)$ and $\gamma(z)$ are step modulation functions for which the system reduces to a non-integrable and non-autonomous model. Usually, the modulation scheme refers to a nonlinearity and dispersion map in the longitudinal direction that may be achieved by periodic concatenation of optical fibers with opposite optical properties. The initial excitation propagates through the optical half-medium with $\beta(z)=1, \gamma(z)=1$,  which reaches $L/2$ where the parameters sharply flip signs to $\beta(z)=-1, \gamma(z)=-1$. Unlike Eqn. \ref{ptnlse_standard} which is integrable, Eq. (5) in general is non-integrable due to the longitudinal modulation scheme, and as such there are no rigorous theoretical methods for predicting or evaluating the exact wave solutions \cite{malomed2006soliton}. In such cases, purely numerical or semi-analytical methods are usually used. 

\begin{figure}
    \centering
    \includegraphics[width=0.99\linewidth]{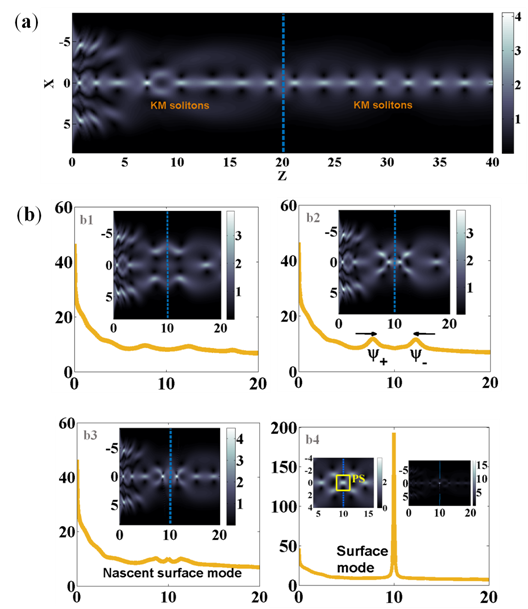}
    \caption{ Evolution of the optical fields in different phases of PT symmetry. (a) Stable KM solitons in the exact PT phase in the absence of transverse shift (Hermitian). Here, $\varepsilon=2.207, \varepsilon_{tsp}=0.0, L=40$. (b) Formation of the surface Peregrine soliton mode at the interface in the broken PT phase. The insets show the spatio-temporal optical field distribution. $(b1) \, \varepsilon=2.02, (b2)\, \varepsilon=2.05, (b3)\, \varepsilon=2.15, (b4)\, $ left inset shows zoom-in view of the surface Peregrine soliton mode encircled by a yellow rectangle ($\varepsilon=2.119 $), right inset shows the subsequent enhanced surface localization ($\varepsilon=2.22$). Here,$ \varepsilon_{tsp}=0.00871, L=20$. The dashed blue line indicates the interface between the distinct optical media (the co-ordinates ($z,x$) have been replaced identically by ($Z,X$) for better visualization.}
    \label{fig2}
\end{figure}

\textit{Emergence of the surface Peregrine soliton}.--The initial Peregrine excitation impinges on the half-space optical medium at $z=z_{0}=0$ and propagates in the media. It gives rise to intriguing wave phenomena in the exact and broken regimes of PT symmetry by nontrivial wave coupling and complex non-Hermitian wave interaction processes. In the absence of any transverse shift in the exact PT phase, stable KM solitons (see Fig. \ref{fig2}(a)) propagate unhindered along the entire composite optical media beyond the optical interface. 

\begin{figure}
    \centering
    \includegraphics[width=0.99\linewidth]{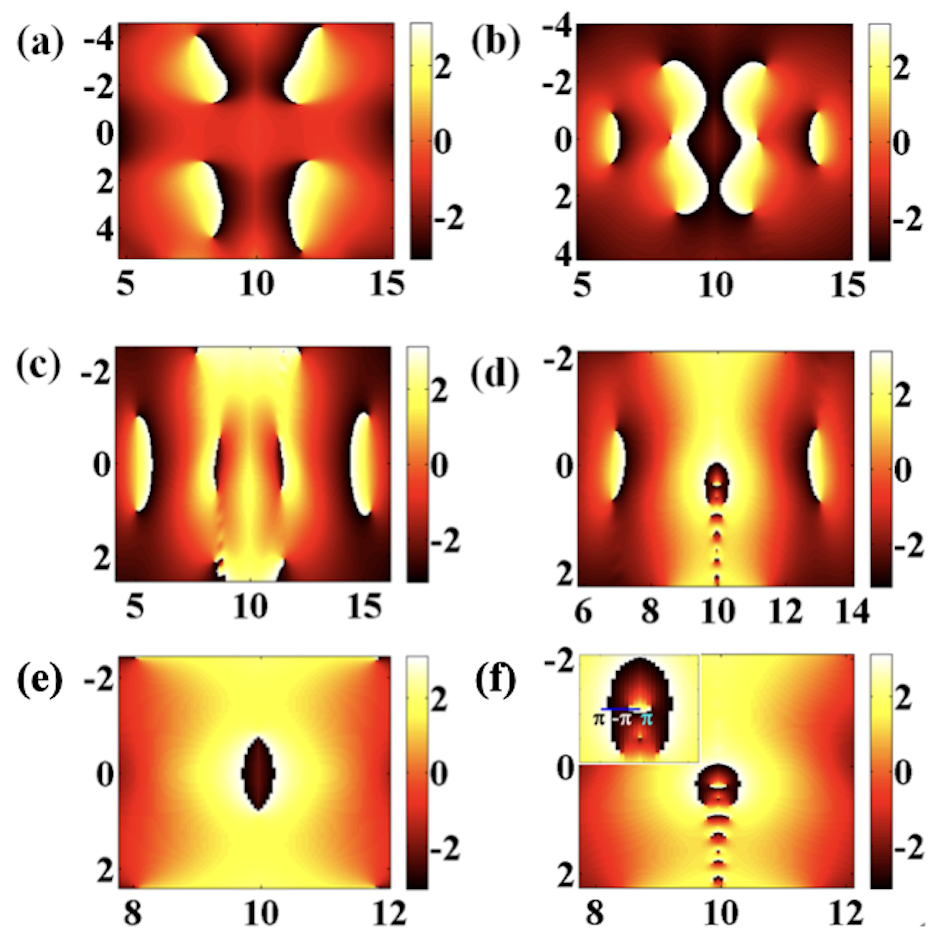}
    \caption{Phase distributions of the optical fields to show the phase jump and the emergence of the Peregrine surface soliton mode. Keeping $\varepsilon_{tsp}=0.00871$, the interval factor parameter is varied as: $(a)\, \varepsilon=2.02, (b)\, \varepsilon=2.10, (c)\, \varepsilon=2.15, (d)\, \varepsilon=2.22$. Taking $\varepsilon=2.22$, we see the effects of transverse shift parameter: (e) $\varepsilon_{tsp}=0.0$ (Hermitian), and (f) $\varepsilon_{tsp}=0.00871$ (non-Hermitian). In the small central portion, Fig. \ref{fig3}(e) reveals a homogeneous phase at a value $-\pi$. In contrast, Fig. \ref{fig3}(f) shows the presence of a sharp phase discontinuity. Inset in Fig. \ref{fig3}(f) shows the central portion with a phase jump across the optical interface.}
    \label{fig3}
\end{figure}

This occurs only when the initial excitation propagates from the optical medium $M_{+,+}$ to $M_{-,-}$. The KM soliton does not propagate beyond the optical interface for other combinations of partnering media as expected due to the broken mirror-reflection symmetry in the wave profiles. The engineering of nonlinearity and spatial dispersion of the optical medium stabilizes the breathing dynamics of the KM soliton beyond the optical interface. It is remarkable to find that the initial excitation propagates stably beyond the optical interface in the PT unbroken regime, despite the fact that the medium is strongly heterogeneous. This stable wave propagation persists even when the number of half-spaces is increased (see Supplementary Information). Only a certain class of wave systems can witness such behaviors \cite{malomed2006soliton}. In stark contrast, enhanced surface wave localization occurs in the broken PT phase via judicious nonlinearity and spatial dispersion engineering of the media, as shown in Fig. \ref{fig2}(b). We argue that this giant surface wave enhancement stems mainly from the interplay between PT breaking of the pseudo-self-induced potential and the longitudinal nonlinearity and spatial dispersion engineering scheme. The interval factor parameter essentially dictates wave coupling and interaction between the initial in-phase two-soliton excitations, resulting in a large enhancement of the optical intensity and accumulation of wave energy at the interface. The two peaks shown in Fig. \ref{fig2} (b (b2)) refer to a pair of self-dual chiral bulk Peregrine and anti-Peregrine solitons in $M_{+,+}$ and $M_{-,-}$. They collide at the interface to form a highly localized surface mode (Fig. \ref{fig2} (b (b4))) followed by a collapse and revival dynamics (Supplementary Information Sec. V). The giant surface amplification is due to the PT breaking of the pseudo-self-induced potential, which is evident from the asymmetric intensity (Fig. 5(d) in the Supplementary Information) and phase distribution (Fig. \ref{fig3}(d) in the main text). In fact, this surface mode is a second-order Peregrine soliton that later evolves into a giant spike-like surface mode. Its origin can be attributed to non-Hermiticity-induced phase distortion (Fig. \ref{fig3} (a-c)), and an abrupt phase transition (Fig. \ref{fig3} (d)) further shown in Fig. \ref{fig3} (e),(f). 

When the initial excitation crosses the interface, the collision of a self-dual pair of Peregrine and anti-Peregrine solitons leads to the surface Peregrine soliton mode, which is enhanced via spontaneous PT breaking of the pseudo-self-induced potential. Unattenuated complete wave tunneling has been demonstrated by excitation of the surface wave at the interface of a balanced gain-loss PT symmetric bilayer system \cite{savoia2014tunneling}. 

\begin{figure}
    \centering
    \includegraphics[width=0.99\linewidth]{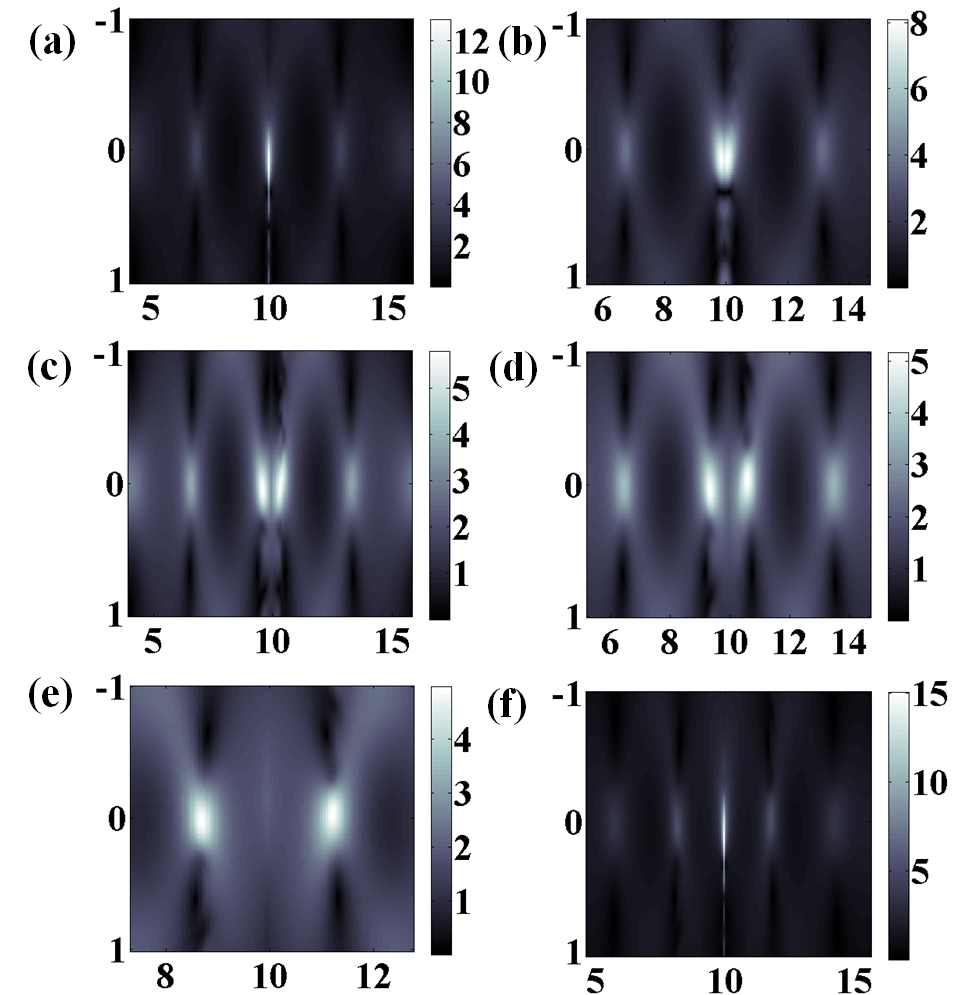}
    \caption{\label{collapse_rev} Collapse and revival dynamics of the surface Peregrine soliton mode at the in the $z-x$ plane as the interval factor is varied: (a) $\varepsilon= 2.2$, (b) $\varepsilon= 2.23$, (c) $\varepsilon= 2.24$, (d) $\varepsilon= 2.25$, (e) $\varepsilon= 2.28$ , (f) $\varepsilon= 2.31$. Here, $\varepsilon_{tsp}=0.00871$.}
    \label{fig4}
\end{figure}

The surface Peregrine soliton mode emerges in a definite parametric regime, and it is sensitive to the initial excitation conditions. The interval factor of the two Peregrine solitons  crucially affects the formation of the surface Peregrine soliton mode in the range of $\varepsilon\in(2.14,2.22)$. Beyond this parameter window, we find the single surface Peregrine soliton mode beginning to split ($\varepsilon=2.23$) into the two localized modes, which recombine again ($\varepsilon=2.28$) at the interface to give rise to the same surface mode before becoming unstable. In this way, they undergo periodic wave \textit{collapse and revival} dynamics (see Fig. \ref{fig4}(a-f)). This periodic collapse and revival dynamics and the emergence of the enhanced surface Peregrine soliton mode at the optical interface are dynamical counterparts of gap closing and re-opening, and the existence of the topological edge state pinned at the topological domain wall \cite{septembre2021parametric, mann2022topological}. The enhanced surface Peregrine soliton mode closely resembles a spike-like extreme event, akin to the fact that a Peregrine soliton is widely known to be a precursor of extreme wave phenomena \cite{kibler2010peregrine,Shrira,gupta2018string}. 

\textit{Associated phase distributions}.--Further numerical analysis confirms that such surface wave phenomena do not exist in the conventional Hermitian analog of NLSE demonstrating its strong non-Hermitian origin. We show the associated phase distributions in the Hermitian and non-Hermitian cases in Fig. \ref{fig3}. The Hermitian case shows an almost homogeneous phase distribution at the center in the $z-x$ plane (Fig. \ref{fig3}(b)), while the non-Hermitian case depicts a distorted phase distribution with phase discontinuity at the interface (Fig. \ref{fig3}(d)). This implies the presence of an abrupt phase transition akin to spontaneous PT breaking of the pseudo-self-induced potential. 

\textit{Topological signatures and non-Hermitian topological phase transition}.--The surface Peregrine soliton reveals underlying topological signatures in certain close topological wave analogies. 

Our model predicts that the two optical half-spaces $M_{+,+}$ and $M_{-,-}$ in the composite optical media contain distinct bulk chiral Peregrine solitons: $\psi(z,x)=(1-\frac{4(1+2iz)}{1+4x^2+4z^2})\,e^{iz}$ in $M_{+,+}$ where $\beta=1$ and $\gamma=1$ and in $M_{-,-}$ $\psi(z,x)=(1-\frac{4(1-2iz)}{1+4x^2+4z^2})\,e^{-iz}$ where $\beta=-1$ and $\gamma=-1$. Interestingly,$\psi_{+}=\psi_{-}^{*}$, which also means that $\hat{P}\hat{T}\psi_{+}=\mathds{1}\psi_{-}$ and $\hat{P}\hat{T}\psi_{-}=\mathds{1}\psi_{+}$. It implies that $\hat{P}\hat{T}$ plays the role of an involution operation under self-duality particle-anti-particle (PA) symmetry. The pair of Peregrine and anti-Peregrine solitons are self-dual to each other. In fact, it is known that a transformation $\bf\Phi$ is self-dual for a set $S=(S_{1},S_{2})$, $\bf{\Phi}$, $(S_{1})$=$ S_{2}$ and $\bf{\Phi}$$(S_{2})$=$ S_{1}$, which exactly corroborates our finding of coexisting PT and self-duality symmetry ($\bf\Phi$=$\hat{P}\hat{T}, \psi_{1,2}\in S_{1,2}$). Self-duality symmetry is known to exist in association with some other kind of symmetry, which in our case turns out to be PT symmetry. Here, $\psi_{\pm}$ refer to the self-dual pair of Peregrine solitons protected by PT symmetry, which collide or annihilate at the optical interface. 

The opposite chiralities of the bulk Peregrine solitons are determined by opposite signs of nonlinearity and spatial dispersion. As predicted by theory, the two bulk chiral Peregrine solitons are found to propagate in opposite directions (Fig. \ref{fig2} (b (b2))) until they collide at the optical interface to give rise to the enhanced surface Peregrine soliton mode. 

It could be worthwhile to note that in the topological insulators, low-energy surface electrons satisfy a Weyl equation, where a full Dirac formalism includes two such equations with opposite handedness. Including a mass term couples the two modes with opposite handedness, and the surface state emerges \cite{stone2015topology}. The surface Peregrine soliton mode appears in close analogy to this topological surface mode. Here, the role of the mass term could be emulated by the parametric engineering of the nonlocal non-Hermitian optical media that couples the two half-spaces with opposite electromagnetic or optical properties. We see that Peregrine and anti-Peregrine solitons exist in the media $M_{+,+}$ and $M_{-,-}$ where both $\beta$ and $\gamma$ are positive and negative, respectively. When certain conditions are met, the surface Peregrine soliton mode appears for the initial excitation propagating from $M_{+,+}$ to $M_{-,-}$. This leads to our conclusion that the surface Peregrine soliton mode originates via an underlying non-Hermitian topological phase transition. 

Alternatively, its topological signature can be understood based on the non-Hermitian photon helicity operator \cite{bliokh2019topological} showing the underlying correspondence between the parameter spaces $\varepsilon-\mu$ (Maxwell EM theory) and $\beta-\gamma$ (our model), or based on the genus and the number of oscillating phases \cite{marcucci2019topological}. Nontrivial topological features can be induced solely by non-Hermiticity \cite{gu2021controlling,gao2020observation,liu2020gain,savoia2014tunneling}. Here, PT breaking of the pseudo-self-induced potential leads to the formation of a topological surface Peregrine soliton mode in the otherwise topologically trivial Hermitian bulk media. These indicate the non-Hermitian topological origin of the surface Peregrine soliton mode at the crossroads of nonlinearity, topology, and non-Hermitian singular processes. 

\textit{Conclusion}.-- The emergence of the surface Peregrine soliton mode is exhibited in a nonlocal PT symmetric nonlinear Schrödinger system with nonlinearity and spatial dispersion-engineered optical media. In particular, we show the stable propagation of the soliton in the exact PT phase and the existence of a space-time-localized surface Peregrine soliton mode in the broken PT phase at the interface between two distinct strongly heterogeneous optical media. We argue that such a surface Peregrine soliton mode appears via enhanced nonlinearity owing to spontaneous breaking of the nonlocal PT symmetric pseudo-self-induced potential and the collision of a self-dual pair of Peregrine and anti-Peregrine solitons. 

Depending on parametric regimes, both stable propagative solutions and enhanced surface localization can be induced in the same physical system. The propagation of stable soliton through this strongly heterogeneous medium could be a robust information carrier, and the enhanced surface localization could be indicative of the trapping of optical or EM energy and sensing at the interface. The existence of such complex wave phenomena thus may have interesting applications in wave manipulation in integrated optical and complex scattering media, especially when endowed with topological protection.

\begin{acknowledgments}
\textit{Acknowledgments}.  S. K. G. thanks A. K. Sarma, M.-H. Lu, and V. V. Konotop for discussions. S.K.G. acknowledges support through a research position at CFTC, Universidade de Lisboa. S. K. G. dedicates this work in memory of his mother, Sumitra Gupta (1957-2021), a science and education enthusiast. He fondly remembers how she would show him the glowing mystic light of a full moon and the host of sunshine flickering upon a river. \\

\end{acknowledgments}

\bibliography{refs}

\end{document}